\documentclass[11pt]{article}
\usepackage{amsmath}
\usepackage{amssymb}
\usepackage[dvips]{epsfig}

 \advance \textwidth by -2in
 \advance \textheight by -3in
\def\##1{{\underline #1}}
\def\*#1{\tilde{{\underline #1}}}
\def\=#1{\underline{\underline{#1}}}
\def\+#1{\tilde{\underline{\underline{#1}}}}

\def\le{\left(}
\def\ri{\right)}
\def\les{\left[}
\def\ris{\right]}

\def\c#1{\cite{#1}}

\def\r#1{(\ref{#1})}

\def\.{\mbox{ \tiny{$^\bullet$} }}
\def\,{\thinspace}

\def\rt{\le \#x,t\ri}
\def\ro{\le \#x,\omega\ri}
\def\partialt{\frac{\partial}{\partial t}\,}
\def\curl{\nabla\times}
\def\div{\nabla\.}

\def\epso{\epsilon_{\scriptscriptstyle 0}}

\def\muo{\mu_{\scriptscriptstyle 0}}

\def\eps{\epsilon}

\begin{document}

\vskip 0.4cm

\begin{center}
{\large {\bf Boundary value problems  and the validity of
the Post constraint in modern electromagnetism}}
\vskip 0.2cm

\noindent  {Akhlesh Lakhtakia}\footnote{Tel: +1 814 863 4319; Fax: +1 814 865 9974;
E--mail: akhlesh@psu.edu}
\vskip 0.2cm
\noindent {\em Computational \& Theoretical Materials Sciences Group (CATMAS)\\
Department of Engineering Science \& Mechanics\\
Pennsylvania State University, University Park, PA 16802--6812, USA}\\
\medskip
and\\
\medskip
\noindent{\em Department of Physics, Imperial College, London SW7 2AZ, United Kingdom}

\end{center}

\noindent {\bf Abstract:} When a (frequency--domain) boundary value problem involving
a homogeneous linear material is solved
to assess the validity of the Post constraint, a conflict arises between the
fundamental differential equations of electromagnetism in the chosen material
and a na\"ive application
of  the usual boundary conditions. It is shown here that the conflict vanishes when the
boundary conditions are properly derived from the fundamental equations, and the
validity of the Post constraint in modern macroscopic electromagnetism
is thereby reaffirmed.

\vskip 0.2cm
\noindent {\em Keywords:\/} Boundary conditions; Electromagnetic theories;  Linear materials;
Magnetoelectric
materials;  Post constraint; Tellegen parameter

\section{Introduction}
The genesis of the Post constraint on the electromagnetic constitutive relations of linear mediums
was described in detail 
quite recently \c{L2004}. This
 {\em structural constraint}  was shown to arise from the following two considerations:
\begin{itemize}
\item Two of the four Maxwell postulates (containing the induction fields and the
sources) should be independent
of the other two Maxwell postulates (containing the primitive fields) at the macroscopic level, just
as the two sets of postulates are mutually independent at the microscopic level.
\item The constitutive functions must be characterized as piecewise
uniform, being born of the spatial homogenization of microscopic entities. Therefore, if
a certain constitutive function
of a homogeneous piece of a medium
cannot be recognized by proper electromagnetic  experimentation, the 
assumption of  a continuously nonhomogeneous analog of that constitutive function
is untenable.
\end{itemize}
Available experimental evidence against the 
validity
of the Post constraint for linear materials was shown to be incomplete and inconclusive, in addition to being based either on the physically inadmissible
premise of purely instantaneous response and/or derived from a pre--modern  version 
of electromagnetism \c{L2004}.

Nevertheless,  solutions of very simple
(frequency--domain) boundary value
problems can be invoked very easily to claim the invalidity of the Post constraint for
linear materials.  Indeed, when a  boundary value problem involving
a homogeneous linear material is formulated
to assess the validity of the Post constraint, a conflict arises between the
fundamental differential equations of electromagnetism in the chosen material
and a na\"ive application
of  the usual boundary conditions. In this paper,
that conflict is  easily resolved~---~in favor of the Post
constraint.

The organization of this paper is as follows: Section \ref{MME} contains a brief review of
modern macroscopic electromagnetism, followed by a relevant presentation of linear
constitutive relations in Section \ref{LCR}. The principal equations of a na\"ive formulation
of boundary value problems are set up in Section \ref{BVP}, and the afore\-mentioned conflict
is presented and resolved in Section \ref{TCR}. The paper concludes with some remarks
in Section \ref{CR}.

\section{Modern Macroscopic Electromagnetism}\label{MME}
Let us begin with the fundamental equations of modern electromagnetism.
The microscopic fields
 are just two: the
electric field $\tilde{\#e}\rt$
and the magnetic field $\tilde{\#b}\rt$.\footnote{The lower--case 
letter signifies that
a field or a source density is microscopic,
while the tilde $\,\tilde{ }\,$
indicates dependence on time. Furthermore,
$\epsilon_o = 8.854\times 10^{-12}$~F/m and
$\mu_o = 4\pi\times 10^{-7}$~H/m are the permittivity 
and the
permeability \index{permeability}
of matter--free space in the absence of an external gravitational field (which
condition is assumed here).} 
These two are accorded the status
of primitive fields in modern electromagnetism, and
their sources are the microscopic charge
density ${\tilde c}\rt$ 
and the microscopic current density
$\tilde{\#j}\rt$.
Both fields and both sources appear in the {\em 
microscopic\/}
Maxwell postulates \c{Jack}
\begin{eqnarray}
\label{eq6.3}
&&\div \tilde{\#e}\rt = \epso^{-1}\,\tilde{c}\rt\,,
\\
\label{eq6.4}
&&\curl\tilde{\#b}\rt -\epso\muo\partialt\tilde{\#e}\rt = 
\muo\,\tilde{\#j}\rt\,,
\\
\label{eq6.5}
&&\div\tilde{\#b}\rt = 0\,,
\\
\label{eq6.6}
&&
\curl\tilde{\#e}\rt +
\partialt\tilde{\#b}\rt = \#0\,.
\end{eqnarray}

Spatial averaging of the microscopic primitive fields and source densities yields
the {\em macroscopic\/}
Maxwell
postulates 
\begin{eqnarray}
\label{eq6.7}
&&\div \tilde{\#E}\rt = \epso^{-1}\,\tilde{\rho}\rt\,,
\\
\label{eq6.8}
&&\curl\tilde{\#B}\rt -\epso\muo\partialt\tilde{\#E}\rt = 
\muo\,\tilde{\#J}\rt\,,
\\
\label{eq6.9}
&&\div\tilde{\#B}\rt = 0\,,
\\
\label{eq6.10}
&&
\curl\tilde{\#E}\rt +
\partialt\tilde{\#B}\rt = \#0\,,
\end{eqnarray}
which involve the macroscopic primitive fields
$\tilde{\#E}\rt$ and $\tilde{\#B}\rt$
as well as
 the macroscopic  source densities $\tilde\rho\rt$
and $\tilde{\#J}\rt$. {\em Equations \r{eq6.7}--\r{eq6.10}
are the fundamental (differential) equations of modern macroscopic electromagnetism.\/}
Let us note that 
\begin{itemize}
\item[(i)] all four equations contain only two fields, both primitive, and 
\item[(ii)]
 all four equations hold 
in matter--free space as well as in matter. 
\end{itemize}
Indeed, modern electromagnetism may be called EB--electromagnetism to indicate the
central role of $\*E\rt$ and $\*B\rt$.

Equations \r{eq6.7}--\r{eq6.10} are not, however, the textbook form of the Maxwell 
postulates. In order
to obtain that familiar form, source densities are decomposed into {\em free\/}
and {\em bound\/} components, and the bound components are then
 quantified through the polarization  and the {magnetization}, both of which are in turn
 subsumed  in the definitions
of the electric induction $\tilde{\#D}\rt$ and the magnetic 
induction $\tilde{\#H}\rt$.
Then, \r{eq6.7}--\r{eq6.10} metamorphose into the following
familiar form:
\begin{eqnarray}
\label{eq6.17}
&&\div \tilde{\#D}\rt = \tilde{\rho}_{so}\rt\,,
\\
\label{eq6.18}
&&\curl\tilde{\#H}\rt -\partialt\tilde{\#D}\rt = 
\tilde{\#J}_{so}\rt\,,
\\
\label{eq6.19}
&&\div\tilde{\#B}\rt = 0\,,
\\
\label{eq6.20}
&&
\curl\tilde{\#E}\rt +
\partialt\tilde{\#B}\rt = \#0\,.
\end{eqnarray}
Here, $\tilde{\rho}_{so}\rt$ and $\tilde{\#J}_{so}\rt$ represent free or externally impressed source
densities.
Let us note that $\tilde{\#H}\rt$ and $\tilde{\#D}\rt$ do not have microscopic
counterparts and therefore are not considered fundamental in modern electromagnetism.

\section{Linear Constitutive Relations}\label{LCR}
The most general
linear constitutive relations may be written as \c{L2004}
\begin{eqnarray}
\nonumber
&&\*D\rt = \int\int \  \+\eps({\#x},t;{\#x}_h,t_h)\.\*E({\#x}-{\#x}_h,t-t_h) \,
d{\#x}_h\, dt_h
\\[5pt]
\nonumber
& &\qquad+\int\int \  \+\alpha({\#x},t;{\#x}_h,t_h)
\.\*B({\#x}-{\#x}_h,t-t_h) \,
d{\#x}_h \,dt_h \, 
\\[5pt]
\label{dconrel1}
& &\qquad+\int\int \  
{\tilde\Phi}({\#x},t;{\#x}_h,t_h)
\,\*B({\#x}-{\#x}_h,t-t_h) \,
d{\#x}_h \,dt_h \, 
\end{eqnarray}
and
\begin{eqnarray}
\nonumber
&&\*H\rt = \int\int \  \+\beta({\#x},t;{\#x}_h,t_h) 
\.\*E({\#x}-{\#x}_h,t-t_h) \,
d{\#x}_h\, dt_h
\\[5pt]
\nonumber
&&\qquad+\int\int \  \+\nu({\#x},t;{\#x}_h,t_h)\.\*B({\#x}-{\#x}_h,t-t_h) \,
d{\#x}_h\, dt_h \, 
\\[5pt]
\label{hconrel1}
&&\qquad -
\int\int \  {\tilde\Phi}({\#x},t;{\#x}_h,t_h)
\,\*E({\#x}-{\#x}_h,t-t_h) \,
d{\#x}_h\, dt_h
\end{eqnarray}
wherein the integrals extend only over the causal values of
$({\#x}_h,t_h)$ in relation to $\rt$. Five  constitutive functions are present in the two
foregoing equations:
$\+\eps$ is the permittivity tensor;
$\+\nu$ is the impermeability tensor;
$\+\alpha$  and
$\+\beta$ are
the magnetoelectric tensors such that
\begin{equation}
{\rm Trace}\les\+\alpha({\#x},t;{\#x}_h,t_h) -\+\beta({\#x},t;{\#x}_h,t_h) \ris \equiv 0\,;
\end{equation}
and ${\tilde \Phi}$ may be called the Tellegen parameter.

When \r{dconrel1} and \r{hconrel1}
are substituted in \r{eq6.17}--\r{eq6.20} to retain only the primitive fields and the 
source densities, the resulting four equations contain
 $\+\eps$, $\+\alpha$, $\+\beta$ and $\+\nu$  
in two ways: 
\begin{itemize}
\item[(i)] by themselves, and 
\item[(ii)] through their space-- and time--derivatives. 
\end{itemize}
In
contrast,  ${\tilde\Phi}$ does not occur by itself, but only in terms of derivatives \c{L2004}.
The  elimination of this anomalous situation leads to the Post constraint
\begin{equation}
\label{Post6.18}
{\tilde\Phi}({\#x},t;{\#x}_h,t_h)\equiv 0\,.
\end{equation}

Arguments in favor of and against the Post constraint were cataloged some years ago \c{WL98},
with the opposing arguments based on the so--called EH electromagnetism wherein  $\*H\rt$ is
regarded as
the primitive magnetic field  and $\*B\rt$ as the induction magnetic field. The EH--electromagnetism is a pre--modern formulation that is still widely used in frequency--domain
research. Opposing arguments
of a similar nature have also been made under the rubric of the heterodox
EDBH--electromagnetism \c{HO-1}, wherein $\*D\rt$ and $\*H\rt$ are also supposed to have microscopic
counterparts and are therefore also considered primitive.

\section{Boundary Value Problems}\label{BVP}

Constitutive functions are macroscopic entities arising from the homogenization 
of assemblies of microscopic bound source densities, with matter--free 
space serving as the reference medium \c{W03}.
In any small enough portion of matter that is homogenizable, the constitutive functions are
uniform. When such a portion will be interrogated for characterization, it will have to be embedded
in matter--free space. Typically, macroscopically homogeneous matter is characterized in the
frequency domain. Hence, it is sensible to investigate if the Tellegen parameter can
be determined by such a measurement.

Without loss of generality, let us consider therefore that all space is divided into two regions, $V_{+}$ and
$V_{-}$, separated by a boundary $S$.
 The region $V_{+}$ is not filled with matter, whereas the region $V_{-}$
is filled with a spatially homogeneous, temporally invariant and spatially local 
matter characterized by the constitutive relations
\begin{eqnarray}
\nonumber
&&
\left.\begin{array}{ll}
\#D\ro = \=\eps(\omega)\.\#E\ro + \=\alpha(\omega)\.\#B\ro + \Phi(\omega)\,\#B\ro\\[4pt]
\#H\ro = \=\beta(\omega)\.\#E\ro + \=\nu(\omega)\.\#B\ro-\Phi(\omega)\,\#E\ro
\end{array}\right\},
\\[5pt]
&&\qquad \qquad{\#x}\in V_{-}\,,
\label{new1}
\end{eqnarray}
where $\omega$ is the angular frequency, and $\#D\ro$ is the temporal Fourier transform
of $\*D\rt$, etc. 

The frequency--domain  differential equations
\begin{equation}
\left.\begin{array}{l}
\div\#B\ro=0\\[5pt]
\curl\#E\ro -
i\omega{\#B}\ro = \#0
\end{array}\right\}\,,\quad \#x\in V_+\cup V_-,
\label{new4}
\end{equation}
are applicable in both $V_+$ and $V_-$, with $i=\sqrt{-1}$.

The
remaining two Maxwell postulates in matter--free space may be written as
\begin{equation}
\left.\begin{array}{l}
\epso\div\#E\ro=\rho_{so}\ro\\[5pt]
\muo^{-1}\curl{\#B}\ro +i\omega\epso{\#E}\ro = 
{\#J}_{so}\ro
\end{array}\right\}\,,\quad \#x\in V_+\,,
\label{anew1}
\end{equation}
in terms of only the macroscopic primitive fields, with
sources that are sufficiently removed from the boundary $S$
\c{Staffan}.
The fields $\#E\ro$ and $\#B\ro$ in $V_{+}$ can be represented using standard techniques \c{Jack,BSU}, and the representations of  $\#D\ro=\epso\#E\ro$ and $\#H\ro=\muo^{-1}\#B\ro$ in $V_{+}$ then follow.

In $V_-$, the  remaining two Maxwell postulates are expressed as follows:
\begin{equation}
\left.\begin{array}{l}
\div\#D\ro=0\\[5pt]
\curl{\#H}\ro +i\omega {\#D}\ro = 
\#0
\end{array}\right\}\,,\quad \#x\in V_-\,,
\label{bnew1}
\end{equation}
Substituting \r{new1} therein, we obtain
\begin{eqnarray}
\nonumber
&&
\div\les
\=\eps(\omega)\.\#E\ro + \=\alpha(\omega)\.\#B\ro\ris \\[5pt]
&&\qquad+\, \Phi(\omega)\,
\div\#B\ro=0\,,\quad \#x\in V_-\,,
\end{eqnarray}
and
\begin{eqnarray}
\nonumber
&&
\curl\les{\=\beta(\omega)\.\#E}\ro\ris +i\omega\=\eps(\omega)\. {\#E}\ro\\[5pt]
\nonumber
&&\qquad +\,
\curl\les{\=\nu(\omega)\.\#B}\ro\ris +i\omega\=\alpha(\omega)\. {\#B}\ro \\[5pt]
&& \qquad\qquad
-\,\Phi(\omega)\les\curl{\#E}\ro -i\omega {\#B}\ro \ris
= 
\#0\,,\quad \#x\in V_-\,.
\end{eqnarray}
These equations simplify to
\begin{equation}
\div\les
\=\eps(\omega)\.\#E\ro + \=\alpha(\omega)\.\#B\ro\ris  =0\,,\quad \#x\in V_-\,,
\label{anew2}
\end{equation}
and
\begin{eqnarray}
\nonumber
&&
\curl\les{\=\beta(\omega)\.\#E}\ro\ris +i\omega\=\eps(\omega)\. {\#E}\ro\\[5pt]
&&\qquad +\,
\curl\les{\=\nu(\omega)\.\#B}\ro\ris +i\omega\=\alpha(\omega)\. {\#B}\ro  = 
\#0\,,\quad \#x\in V_-\,,
\label{new2}
\end{eqnarray}
by virtue of \r{new4}.
For many classes of materials and shapes of $S$, $\#E\ro$ and $\#B\ro$ in $V_{-}$ can also be adequately
represented \c{Belbook,Joshua}; and thereafter so can be $\#D\ro$ and $\#H\ro$ in $V_{-}$.

In order to solve the boundary value problem, the boundary conditions 
\begin{equation}
\left.\begin{array}{ll}
\#B^{norm}({\#x}+,\omega) =\#B^{norm}({\#x}-,\omega) \\[5pt]
\#D^{norm}({\#x}+,\omega) =\#D^{norm}({\#x}-,\omega) \\[5pt]
\#E^{tan}({\#x}+,\omega) = \#E^{tan}({\#x}-,\omega)\\[5pt]
\#H^{tan}({\#x}+,\omega) = \#H^{tan}({\#x}-,\omega)
\end{array}\right\}\,, \quad \#x\in S\,,
\label{new3}
\end{equation}
have to be imposed
on the boundary $S$.  Here,  $\#B^{norm}({\#x}\pm,\omega)$
indicate the normal components of $\#B\ro$ on either side of $S$, 
whereas  $\#E^{tan}({\#x}\pm,\omega)$
denote the tangential components of $\#E\ro$ similarly, etc.
Some resulting set of equations can then be solved to determine the scattering of an incident
field by the material contained in $V_-$.

Much effort is not required to solve the simplest boundary value problems. Relevant to the Post constraint,
reference is made to two papers wherein the boundary $S$ is 
a specularly smooth plane of
infinite extent \c{LD91,LSV91}. More complicated boundaries have also been tackled
 \c{Joshua,Mon90,Lak92}. The inescapable conclusion from examining  the results
 of boundary value problems is that the fields
 scattered in $V_+$ by the material contained in $V_-$ are affected by the Tellegen
 parameter (if any). Yet that conclusion is na\"ive and incorrect, as we see next.
 
 \section{The Conflict and Its Resolution}\label{TCR}
 We have two very sharply contrasting Statements
 emanating from the foregoing frequency--domain exercise:
 \begin{itemize}
 \item[A.] The Tellegen parameter $\Psi$ vanishes from the fundamental
 equations \r{new4}, \r{anew2} and \r{new2} for the material of which the chosen scatterer is made.
 \item[B.] The fields scattered by the chosen scatterer contain a signature of
 the Tellegen parameter (if any).
 \end{itemize}
 In other words, the Tellegen parameter is a {\em ghost}: it does not have a
 direct existence in the fundamental differential equations, but its presence may be indirectly
 gleaned from a scattering measurement.
 
 The ghostly nature of the Tellegen parameter is a consequence of the
 boundary conditions \r{new3}$_2$ and \r{new3}$_4$. Even more specifically, it arises from
 the representations of $\#D\ro$
 and $\#H\ro$ in $V_-$. It is instructive to decompose the macroscopic induction fields
 as \c{Lsst}
 \begin{equation}
 \left.\begin{array}{l}
 \#D\ro = \#D_{actual}\ro + \#D_{excess}\ro\\[5pt]
 \#H\ro = \#H_{actual}\ro + \#H_{excess}\ro
 \end{array}\right\}\,, \quad \#x\in V_-\,,
 \end{equation}
 where
 \begin{equation}
  \left.\begin{array}{l}
   \#D_{actual}\ro=  \=\eps(\omega)\.\#E\ro + \=\alpha(\omega)\.\#B\ro\\[5pt]
 \#H_{actual}\ro=  \=\beta(\omega)\.\#E\ro + \=\nu(\omega)\.\#B\ro
 \end{array}\right\}\,, \quad \#x\in V_-\,,
 \end{equation}
 are retained in  \r{anew2} and \r{new2}.
 On the other hand,
 \begin{equation}
   \left.\begin{array}{l}
     \#D_{excess}\ro = \Phi(\omega)\,\#B\ro\\[5pt]
  \#H_{excess}\ro = -\Phi(\omega)\,\#E\ro
  \end{array}\right\}\,,\quad \#x\in V_-\,,
  \end{equation}
are filtered out of \r{anew2} and  \r{new2} by 
 \r{new4}  but do affect the boundary conditions \r{new3}$_2$ and \r{new3}$_4$.

The fundamental differential equations in $V_-$ can now be written
as follows:
\begin{equation}
\left.\begin{array}{l}
\div\#B\ro=0\\[5pt]
\curl\#E\ro -
i\omega{\#B}\ro = \#0\\[5pt]
\div\#D_{actual}\ro = 0\\[5pt]
\curl\#H_{actual}\ro 
+i\omega\#D_{actual}\ro  = 
\#0
\end{array}\right\}
\,,\quad \#x\in V_-\,.
\label{new5}
\end{equation}
Boundary conditions in electromagnetics emerge from the fundamental
equations \c{JZB85}. Therefore, consistently with \r{new5}, the
{\em correct\/} boundary conditions on $S$ are
\begin{equation}
\left.\begin{array}{ll}
\#B^{norm}({\#x}+,\omega) =\#B^{norm}({\#x}-,\omega) \\[5pt]
\#D^{norm}({\#x}+,\omega) =\#D^{norm}_{actual}({\#x}-,\omega) \\[5pt]
\#E^{tan}({\#x}+,\omega) = \#E^{tan}({\#x}-,\omega)\\[5pt]
\#H^{tan}({\#x}+,\omega) = \#H^{tan}_{actual}({\#x}-,\omega)
\end{array}\right\}\,, \quad \#x\in S\,,
\label{new6}
\end{equation}
instead of \r{new3}. 
Thus the correct formulation of the boundary value problem involves \r{new6}$_2$ and
\r{new6}$_4$
instead of  \r{new3}$_2$ and \r{new3}$_4$.

To sum up, the conflict between Statements A and B arises from a na\"ive and incorrect formulation
of the boundary value problem. The correct formulation does not
contain   $\#D_{excess}\ro$
and $\#H_{excess}\ro$ in $V_-$ as well as in the boundary conditions.

\section{Concluding Remarks}\label{CR}

Any field that cannot survive in the fundamental differential equations is superfluous.
Neither $  \#H_{excess}\ro$ nor $  \#D_{excess}\ro$ survives, and may therefore
be discarded {\em ab initio\/}. The Post constraint thus removes the nonuniqueness 
inherent in \r{new1}, not to mention
 in  \r{dconrel1} and \r{hconrel1}, which can appear in two of the four Maxwell
 postulates in relation to the other two 
 postulates. No wonder, de Lange and Raab \c{deLR1,deLR2} could recently
complete a major exercise~---~whereby a multipole formulation of linear materials
that was initially noncompliant with the Post constraint was made compliant.  

In addition, the Post constraint also removes two anomalies: the first is that of a constitutive
function not appearing by itself but only through its derivatives \c{L2004}; the
second is that of the Tellegen ``medium''  which is isotropic (i.e., with  
direction--independent properties) but wherein propagation characteristics
in antiparallel directions are different.

A simple exercise shows that isolated magnetic monopoles can negate the validity
of the Post constraint  \c{Booj,Dmit}, but
the prospects of observing such a magnetic monopole are rather remote \c{JL95,Hagi}.
Furthermore, although the electromagnetic characterization
of matter--free space, even in the context of general relativity, is compliant with the
Post constraint \c{Plebanski},  the axion concept renders that constraint invalid \c{HO-1}.
No axions have yet been detected however \c{Phys_Today_May_2005}.  Finally, available data on
magnetoelectric materials seems to negate the Post constraint
\c{Dzya,FRS,ODell}, but that data is faulty \c{L2004}
as it is based on the neglect of causality \c{Linv} and a false manipulation
of the Onsager principle \c{LD2005}. Needless to add, if either an isolated magnetic monopole or
an axion is ever discovered, or if a magnetoelectric material is properly characterized to have
the electromagnetic properties claimed for it by virtue of misapplications of various
principles, the Post constraint would be invalidated and the basics of EB--electromagnetism
would have to thought anew.

\bigskip
\noindent {\bf Acknowledgment.} Occasional discussions with Dr. E.J. Post are gratefully acknowledged.
Thanks are also due to the Department of Management Communication, University of
Waikato, Hamilton, New Zealand, for hospitality during a visit when this paper
was written.


\end{document}